\documentclass[conference]{IEEEtran}
\IEEEoverridecommandlockouts
\usepackage{booktabs}
\usepackage{multirow}
\usepackage{amsmath,amssymb,amsfonts}
\usepackage{graphicx}
\usepackage{cite}
\ifCLASSINFOpdf
\else
\fi
\begin{document}
%
\title{Bayesian PET Reconstruction with Learned Flow Matching Priors via Langevin Sampling}

\author{\IEEEauthorblockN{Hengjia Ran}
\IEEEauthorblockA{ZJU-UIUC Institute\\
College of Information Science and\\ Electronic Engineering\\
Zhejiang University, China\\
}
\and
\IEEEauthorblockN{Huafeng Liu}
\IEEEauthorblockA{College of Biomedical Engineering\\
and Instrument Science\\
Zhejiang University, China\\
}
\and
\IEEEauthorblockN{Bo Zhao$^{\dagger}$
\thanks{$\dagger$\text{ Corresponding author: Bo Zhao (bozhao@intl.zju.edu.cn).}}
}
\IEEEauthorblockA{ZJU-UIUC Institute\\
College of Information Science and\\ Electronic Engineering\\
Zhejiang University, China}}


%


\maketitle

\begin{abstract}
Image reconstruction for positron emission tomography (PET) is a challenging inverse problem due to its inherent ill-posedness and Poisson counting noise, particularly in low-dose acquisitions. In this work, we introduce a novel Bayesian PET reconstruction framework that leverages a flow-matching generative prior learned from high-quality PET images. By incorporating this prior, we formulate PET reconstruction as a Bayesian inference problem and solve it via latent-space posterior sampling using the
unadjusted Langevin algorithm. We evaluate the proposed method through numerical experiments and benchmark it against classical reconstruction methods and state-of-the-art deep learning–based baselines. The proposed method achieves improved reconstruction performance while enabling posterior uncertainty quantification.

\textit{Index Terms\textemdash}Flow matching, generative models, Bayesian inference, posterior sampling, positron emission tomography.
\end{abstract}

%
\IEEEpeerreviewmaketitle

\section{Introduction}
Positron emission tomography (PET) is a powerful molecular imaging technique that provides unique insights into biological processes and metabolism \cite{cherry2012physics, valk2006pet}. However, reconstructing high-quality PET images from measured sinogram data remains a significant challenge due to the inherent ill-posedness of the inverse problem and the presence of Poisson counting noise. These issues are particularly pronounced in low-dose acquisitions designed to reduce patient exposure to ionizing radiation.

Analytic reconstruction methods, such as filtered backprojection \cite{wernick2004emission}, have been applied to PET reconstruction. However, these methods typically rely on simplified imaging models and are therefore prone to noise amplification and artifacts. Statistical iterative reconstruction methods, such as maximum-likelihood (ML) approaches \cite{shepp1982maximum, hudson1994accelerated}, improve upon analytic methods by incorporating more accurate models of imaging systems and noise statistics. Nonetheless, due to the ill-posed nature of the underlying inverse problem, they can still suffer from substantial noise contamination, particularly in low-dose acquisitions. To address this issue, various penalized maximum-likelihood (PML) and maximum a posteriori (MAP) reconstruction methods \cite{Levitan1987MAPEM, hebert1989gem, green1990bayesian, fessler1995sage, Zhou2007Wavelet, burger2014total, wang2015pet, chen2015sparse} have been introduced by incorporating hand-crafted priors. Although these methods can improve the signal-to-noise ratio (SNR), they may introduce blocky artifacts and oversmooth anatomical structures.

Over the past decade, deep learning–based reconstruction has emerged as a powerful paradigm for PET imaging \cite{Reader2021DeepLearning}. Supervised learning-based methods, including end-to-end reconstruction \cite{haggstrom2019deeppet,hu2021dpir} and deep unrolling \cite{hu2023dynamic,gong2019emnet,mehranian2020model}, have demonstrated promising performance by learning from paired training data. However, they often require extensive labeled datasets and may have limited generalization capability. In contrast, unsupervised learning-based methods leverage inherent image-structure information—such as the inductive bias of network architectures \cite {gong2018pet} or implicit continuity \cite {moussaoui2025implicit}—thereby alleviating the need for paired training data. Nonetheless, their performance is often constrained by limited model expressiveness. Recently, generative models\cite{ho2020ddpm, song2021score_sde}, especially score-based diffusion models, have demonstrated effectiveness in learning expressive image prior distributions from training data. Although promising results have been obtained in PET reconstruction \cite {singh2024score, Webber2025LikelihoodScheduled}, these methods often rely on sophisticated mechanisms to maintain data consistency and incur high computational costs due to iterative multi-step sampling.

In this work, we introduce a novel Bayesian PET reconstruction method based on posterior sampling guided by a flow-matching generative prior \cite{lipman2023flow, liu2022flow}. Flow matching models a continuous-time probability flow via a neural ordinary differential equation (ODE), providing a deterministic and differentiable generative mapping from a simple latent distribution to the target image distribution. Compared with diffusion models \cite{ho2020ddpm, song2021score_sde}, flow matching avoids complex noise schedules and stochastic reverse-time sampling in the generative prior. Specifically, the proposed method first learns an expressive image prior from high-quality PET data and subsequently formulates PET reconstruction as a Bayesian inference problem. To solve the resulting posterior inference problem, we develop a latent-space posterior sampling algorithm based on Langevin dynamics. By integrating the imaging system model and Poisson noise statistics with the learned prior within a principled Bayesian framework, the proposed method achieves improved reconstruction performance while enabling posterior uncertainty quantification. We demonstrate the effectiveness of the proposed method through numerical experiments.

\section{PROPOSED METHOD}
\subsection{Problem Formulation}
PET aims to reconstruct the spatial distribution of the radiotracer concentration $\mathbf{x} \in \mathbb{R}_+^{N}$ from sinogram measurements $\mathbf{y}\in {\mathbb{Z}}_{+}^{M}$, which can be modeled by 
\begin{equation}
\mathbf{y} \sim \text{Poisson} ( \mathbf{A} \mathbf{x} + \mathbf{r} ),
\end{equation}
where $ \mathbf{A}\in {\mathbb{R}}^{M\times N}$ denotes the system matrix, and $\mathbf{r}\in {\mathbb{R}}^{M}$ denotes the expected background events, including scatter and random coincidences. Accordingly, the probability of observing the measurement $\mathbf{y}_i$ in the $i$-th sinogram bin given an image $\mathbf{x}$ can be written as 
 \begin{equation}
p(\mathbf{y}_i|\mathbf{x})=\frac{{({[\mathbf{A}\mathbf{x}]_i+\mathbf{r}_i)}}^{\mathbf{y}_i}{e}^{-([\mathbf{A}\mathbf{x}]_i+\mathbf{r}_i)}}{\mathbf{y}_i!}.
\label{3}
  \end{equation}
Assuming independence across sinogram bins, the log-likelihood function for the entire measurement vector $\mathbf{y}$ is given by
  \begin{equation}
\log p(\mathbf{y}|\mathbf{x})= \sum_{i=1}^{M}\mathbf{y}_{i}\log([{\mathbf{A}\mathbf{x}]}_{i}+\mathbf{r}_{i})-({[\mathbf{A}\mathbf{x}]}_{i}+\mathbf{r}_{i})-\log(\mathbf{y}_i!).
\label{eq:likelihood_fun}
   \end{equation}

Given the above likelihood function that models the underlying imaging system and measurement noise, we can perform ML reconstruction \cite{shepp1982maximum, hudson1994accelerated}. However, due to the ill-posed nature of the inverse problem, ML reconstruction often suffers from significant noise amplification, which compromises its utility. To address this issue, we introduce a Bayesian approach by incorporating an image prior distribution $p(\mathbf{x})$.  

Specifically, we represent the image prior implicitly using a flow-matching generative model \cite{lipman2023flow} trained on a set of high-quality PET images. Flow matching is a state-of-the-art generative modeling framework, which learns a mapping $G_{\theta}(\cdot)$ that transforms a simple base distribution $p_0(\mathbf{x}_0)$ (e.g., a standard normal distribution) to the target distribution $p_1(\mathbf{x}_1)$ via the following ODE: 
\begin{equation}
    \frac{\mathrm{d}\mathbf{x}_t}{\mathrm{d}t} = v_t^{\theta}(\mathbf{x}_t), \quad \text{where} \quad t\in[0,1].
    \label{FM}
\end{equation}
Here ${v}_{t}^{\theta}(\cdot)$ denotes a velocity field parameterized by a neural network with parameters ${\theta}$. We learn $\theta$ by minimizing the following conditional flow matching loss \cite{lipman2023flow}:
\begin{equation}
\mathcal{L}_\text{CFM}(\theta)=\mathbb{E}_{t, \mathbf{x}_1, \mathbf{x}_t}\left[{\left\|{v}_{t}^{\theta}(\mathbf{x}_t)-v_t^\text{target}(\mathbf{x}_t|\mathbf{x}_1)\right\|}^{2}\right],
\label{eq:CFM} 
\end{equation}
where $v_t^\text{target}(\mathbf{x}_t|\mathbf{x}_1)$ denotes the target velocity field, which is the conditional velocity field of the prescribed flow along the conditional probability path $p_t(\mathbf{x}_t|\mathbf{x}_1)$ and the above expectation is taken over $t \sim \text{Unif}{[0,1]}$, $\mathbf{x}_1\sim p_1(\mathbf{x}_1)$, and $\mathbf{x}_t \sim p(\mathbf{x}_t|\mathbf{x}_1)$.

We further adopt the Gaussian conditional optimal transport path, in which intermediate points are linearly interpolated as
\begin{equation}
\mathbf{x}_t=t\mathbf{x}_1+(1-t)\mathbf{x}_0.
\end{equation}
Let $\mathbf{x}_0 \sim \mathcal{N}(\mathbf{0},\mathbf{I})$ and $\mathbf{x}_1\sim p_1(\mathbf{x}_1)$. Then $p_t(\mathbf{x}_t|\mathbf{x}_1)=\mathcal{N}(t\mathbf{x}_1, (1-t)^2\mathbf{I})$. The corresponding target velocity field is 
\begin{equation}
v_t^\text{target}(\mathbf{x}_t|\mathbf{x}_1) = \frac{d{\mathbf{x}}_{t}}{dt} = \mathbf{x}_1 - \mathbf{x}_0.
\end{equation}
Accordingly, \eqref{eq:CFM} can be rewritten as 
\begin{equation}
\mathcal{L}_\text{CFM}(\theta)=\mathbb{E}_{t, \mathbf{x}_1, \mathbf{x}_0}\left[{\left\|{v}_{t}^{\theta}(\mathbf{x}_t)-(\mathbf{x}_1-\mathbf{x}_0)\right\|}^{2}\right],
\end{equation}
where the expectation is taken over $t \sim \text{Unif}{[0,1]}$, $\mathbf{x}_1\sim p_1(\mathbf{x}_1)$, and $\mathbf{x}_0 \sim \mathcal{N}(0,\mathbf{I})$.

After training, we solve the ODE from $t=0$ to $t=1$ as follows:  
\begin{equation}    
\mathbf{x}_1 = \mathbf{x}_0 + \int_0^1 {v}_{t}^{\theta}(\mathbf{x}_t) dt.
\label{eq:ODE_solution}
\end{equation}
The solution in \eqref{eq:ODE_solution} induces a deterministic and differentiable mapping, $ \mathbf{x} = \mathbf{x}_1 = \mathbf{G}_\theta(\mathbf{x}_0)$, which transports an initial latent noise vector $\mathbf{x}_0 \sim \mathcal{N}(\mathbf{0}, \mathbf{I})$ to a sample $\mathbf{x} \sim p(\mathbf{x})$. 
In practice, we solve the ODE numerically, and, in this work, we use the forward Euler method with $\Delta t=1/L$, which yields the following iterative update: 
\begin{equation}
\mathbf{x}_{t+\Delta t} = \mathbf{x}_{t} + \Delta t \cdot {v}_{t}^{\theta}(\mathbf{x}_t).
\end{equation}

With the implicit prior $p(\mathbf{x})$ induced by flow matching, the image reconstruction problem is cast as a Bayesian inference problem with the posterior density formally given by
\begin{equation}
p(\mathbf{x}|\mathbf{y}) = 
\frac{p(\mathbf{x})p(\mathbf{y}|\mathbf{x})}{p(\mathbf{y})} \propto p(y|x)p(x).
\label{eq:Bayesian_inference}
\end{equation}

\begin{figure*}[t]
    \centering
    \includegraphics[width=0.9\linewidth]{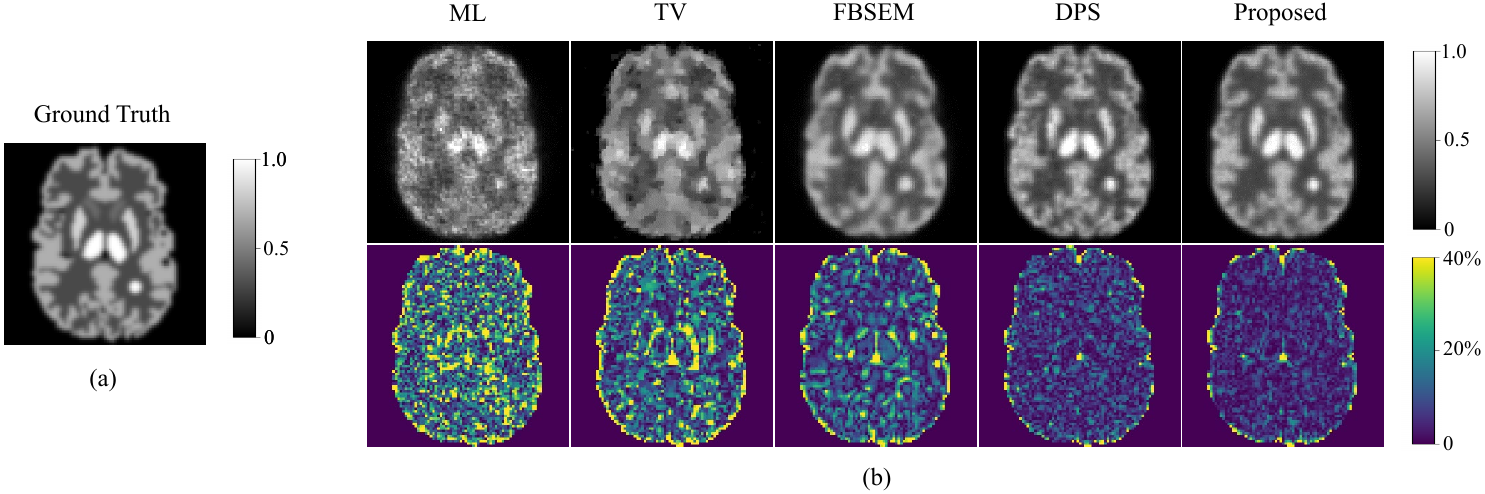} %
    \caption{Comparison of reconstruction methods at the 10\% dose level. (a) Ground-truth. (b) Reconstructions and corresponding error maps for ML, TV, FBSEM, DPS (MMSE estimate), and the proposed method (MMSE estimate). }
    \label{1}
\end{figure*}

 \subsection{Solution Algorithm}
We develop a posterior sampling algorithm to solve the resulting inference problem in \eqref{eq:Bayesian_inference}.
Since the evidence term $p(\mathbf{y})$ in the denominator of \eqref{eq:Bayesian_inference} is intractable, we cannot directly sample $\mathbf{x}$ from $p(\mathbf{x}|\mathbf{y})$. To avoid evaluating the evidence term, we employ gradient-based MCMC methods, which use the gradient of the log-posterior:
\begin{equation}
\nabla_{\mathbf{x}}\log{p(\mathbf{x}|\mathbf{y})} = \nabla_{\mathbf{x}}\log{p(\mathbf{x})} + \nabla_{\mathbf{x}}\log{p(\mathbf{y}|\mathbf{x})}. 
\end{equation}
Several off-the-shelf gradient-based MCMC samplers are available \cite{roberts1996exponential, welling2011sgld, girolami2011riemann, pereyra2016survey, Durmus2017ULA}. In this work, we adopt the unadjusted Langevin algorithm (ULA) \cite{roberts1996exponential, Durmus2017ULA}, which is obtained by discretizing the Langevin diffusion. 

Moreover, for the implicit flow-matching prior, the image-space prior density
$p(x)$ and its score $\nabla_x \log p(x)$ are not directly available. Therefore, we perform posterior sampling in the latent space. After drawing posterior samples in the latent space, we map them to the image space using the generator $G_{\theta}(\cdot)$, i.e., $\mathbf{x} = G_{\theta}(\mathbf{x}_0)$. Specifically, with an initialization $\mathbf{x}_0^{0} \sim \mathcal{N}(\mathbf{0},\mathbf{I})$, the update of the latent-space ULA is given by  
\begin{equation} 
\mathbf{x}_0^{k+1} = \mathbf{x}_0^{k} + {\epsilon} \nabla_{\mathbf{x}_0} \log p(\mathbf{x}_0^{k}|\mathbf{y}) + \sqrt{2\epsilon} \mathbf{w}_k,
\label{eq:ula_update} 
\end{equation} 
where $\epsilon$ is the step size and $\mathbf{w}_k\sim\mathcal{N}(\mathbf{0},\mathbf{I})$ is a random noise vector. 

According to Bayes' theorem, the posterior distribution over $\mathbf{x}_0$ can be written as 
\begin{align}
    \log p(\mathbf{x}_0|\mathbf{y}) = \log p_0(\mathbf{x}_0) + 
    \log p(\mathbf{y}|\mathbf{G}_\theta(\mathbf{x}_0)) + C \nonumber \\
    = -\frac{1}{2}\|\mathbf{x}_0\|_2^2 +
    \sum_{i=1}^{M}  \mathbf{y}_i \log\left([\mathbf{A}\mathbf{G}_\theta(\mathbf{x}_0)]_i + \mathbf{r}_i\right)  \nonumber \\
     - \sum_{i=1}^M\left([\mathbf{A}\mathbf{G}_\theta(\mathbf{x}_0)]_i + \mathbf{r}_i\right)  + C',
    \label{eq:log_noise_space_posterior}
\end{align}
where $C$ and $C'$ are two constants independent of $\mathbf{x}_0$. Consequently, the log-posterior gradient in \eqref{eq:ula_update} is given by 
\begin{equation} 
\nabla_{\mathbf{x}_0} \log p(\mathbf{x}_0|\mathbf{y}) = \mathbf{J}_{G_\theta}(\mathbf{x}_0)^T \mathbf{A}^T \bigg( \frac{\mathbf{y}}{\mathbf{A}G_\theta(\mathbf{x}_0)+\mathbf{r}} - \mathbf{1} \bigg) - \mathbf{x}_0,
\end{equation} 
where the division is element-wise and $\mathbf{1}$ denotes the all-ones vector. Here
$J_{G_\theta}(\mathbf{x}_0)$ denotes the Jacobian of the generator $G_\theta(\cdot)$ evaluated at
$\mathbf{x}_0$. In practice, the corresponding vector-Jacobian product can be efficiently computed by automatic
differentiation without explicitly forming the full Jacobian.

After discarding an initial burn-in period to allow the Markov chain to
approach its stationary regime, we perform Bayesian reconstruction using
posterior image samples obtained by mapping the latent samples through
\(G_\theta(\cdot)\). Moreover, this sampling-based inference approach enables various Bayesian estimators to be formed from the posterior samples. For instance, given posterior samples $\{\mathbf{x}_0^{s}\}_{s=1}^{S}$ in the latent space, we can form the minimum mean-square error (MMSE) estimate of 
$\mathbf{x}$ by 
 \begin{equation} 
 \hat{\mathbf{x}}_{\text{MMSE}} =\frac{1}{S}\displaystyle\sum_{s=1}^{S}G_\theta(\mathbf{x}_0^s). \label{eq:mmse_estimate_fm} 
 \end{equation}
 Additionally, this approach allows for quantification of uncertainty in the reconstruction process. For example, we can calculate the voxel-wise statistics, such as posterior variance, across the reconstructed samples.
 
\begin{figure*}[t]
    \centering
    \includegraphics[width=0.9\linewidth]{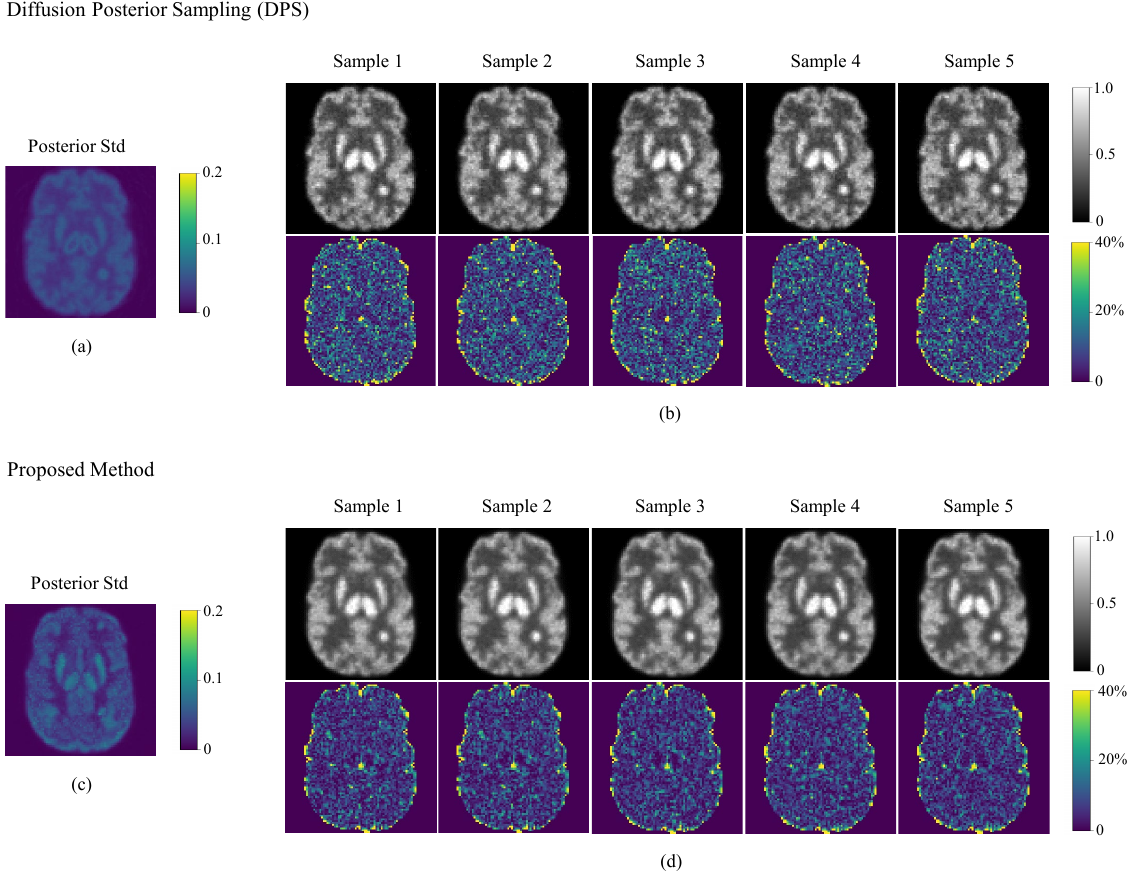} %
    \caption{Reconstruction results and uncertainty visualization at the 10\% dose level: (a) voxel-wise posterior standard deviation estimated by DPS; (b) five representative samples drawn from the DPS posterior, with the corresponding error maps shown below; (c) voxel-wise posterior standard deviation estimated by the proposed ULA-based method; (d) five representative samples drawn from the ULA posterior, with the corresponding error maps shown below. It is clear that the proposed method produces posterior samples with lower error and reduced artifacts.}
    \label{2}
\end{figure*}

\section{Numerical Results}
\label{sec:pagestyle}
\subsection{Experiment Setup}
\noindent{\bf Dataset.} A realistic 3D digital brain phantom ($128 \times 128 \times 60$ voxels, 2 mm isotropic voxel size), derived from the Zubal anthropomorphic phantom \cite{zubal1994computerized}, was used for simulation studies. The phantom was augmented with three spherical lesions (radii of 1.5–2.5 voxels) exhibiting elevated radiotracer uptake. Time-activity curves were simulated using a two-tissue compartment model, with kinetic parameters randomly sampled from normal distributions (with a coefficient of variation of 30\%) to introduce biological variability across voxels. The 10th frame (5 min post-injection) was selected to represent a static PET acquisition. Scanner geometry was modeled after a Siemens Biograph mCT \cite{noh2009statistical}, with sinograms comprising 256 radial bins and 224 projection angles. A spatially-invariant Gaussian point spread function (PSF) with 4.5 mm full-width at half-maximum (FWHM) was applied. Attenuation and detector sensitivity effects were incorporated using precomputed correction factors for each line of response. 

A low-count acquisition corresponding to 10\% of the full dose was generated by scaling the number of true coincidences from a reference full-dose acquisition containing \(4\times 10^6\) counts. Additional uniformly distributed random and scatter events, corresponding to 30\% of the true coincidences, were added to all simulations.

To train the prior, a dataset of 12,000 high-quality 2D slices was generated by reconstructing full-dose data using the EM algorithm. The same dataset was used to train the supervised baseline. Evaluation was performed on a separate test set of 2,000 images.

\noindent{\bf Implementation details.} The velocity field ${v}_{t}^{\theta}(\mathbf{x}_t)$ for flow matching was represented using a four-layer U-Net \cite{ronneberger2015unet}, and was trained using the Adam optimizer \cite{kingma2014adam} with an initial learning rate of $1\times10^{-4}$. During reconstruction, we solved the ODE using the forward-Euler method with 20 time steps. 
For ULA sampling, we set the step size $\epsilon=10^{-3}$ and employed a total number of $T$ = 150,000 iterations, including a burn-in period of $T_\text{burn} = 50,000$ and a sampling period of $T_\text{sample} = 100,000$. All experiments were conducted on a Linux server equipped with two NVLink-connected NVIDIA H100 NVL GPUs (each with 96 GB of HBM3 memory) and 1.5 TB of DDR5 RAM, running PyTorch 2.8.0 and Python 3.10. 

\noindent{\bf Method Comparison.} 
The proposed method was compared with: 1) ML reconstruction \cite{shepp1982maximum}; 2) TV reconstruction\cite{burger2014total}; 3) deep unrolling reconstruction (FBSEM\cite{mehranian2020model}); 4) score-based diffusion posterior sampling reconstruction (DPS\cite{singh2024score, chung2022diffusion}). To facilitate a fair comparison with the diffusion-based reconstruction method, we used the same U-Net
architecture and trained the diffusion model for 3 hours to ensure convergence and optimize performance. During inference, the total sampling time of the proposed method
(5 hours for 150,000 samples, including 50,000 burn-in samples) was matched to that of DPS, which generated 800 samples within the same time budget. Quantitative comparisons were performed using the MMSE estimates computed from the corresponding posterior sample sets. We evaluated all methods using three quantitative metrics: normalized root mean square error (NRMSE), peak signal-to-noise ratio (PSNR), and structural similarity index
(SSIM) \cite{Wang2004SSIM}.

\begin{table}[t]
\centering
\caption{Quantitative comparison of reconstruction methods at the 10\% dose level. The best results are highlighted in \textbf{bold}.}
\label{tab:quantitative_comparison_revised}
\normalsize
\begin{tabular}{|c|c|c|c|c|} 
\hline 
\# & Method   & NRMSE ($\downarrow$) & PSNR ($\uparrow$) & SSIM ($\uparrow$) \\
\hline 
1 & ML       & 20.46\%           & 26.14           & 0.861           \\
2 & TV       & 17.09\%           & 27.71           & 0.876           \\
3 & FBSEM    & 13.32\%           & 29.87          & 0.922 \\
4 & DPS      & 8.29\%           & 33.99         & 0.981 \\
5 & Proposed & \textbf{7.35\%}   & \textbf{35.04}  & \textbf{0.983}  \\
\hline 
\end{tabular}
\end{table}

\subsection{Results}
Figure~\ref{1} shows reconstructed images and corresponding error maps obtained using ML, TV, FBSEM, DPS, and the proposed method at the 10\% dose level. Table~\ref{tab:quantitative_comparison_revised} summarizes the quantitative results of these methods. As can be seen, the ML reconstruction is severely corrupted by noise and artifacts. TV reconstruction improves upon the ML reconstruction, but suffers from excessive smoothing and edge artifacts. In contrast, the three deep learning–based methods demonstrate significantly better performance at this low-dose acquisition. Notably, the proposed method outperforms FBSEM and DPS both qualitatively and quantitatively.

Figure 2 further compares posterior sampling results obtained by DPS and the proposed method. The proposed method produces posterior samples with visibly fewer artifacts and
lower reconstruction error than DPS. In addition, the estimated posterior standard deviation provides voxel-wise uncertainty information, demonstrating the capability of the proposed Bayesian framework to quantify reconstruction uncertainty.

\section{Conclusion}
\label{sec:conclusion}
In this paper, we presented a Bayesian PET reconstruction method that leverages a learned flow-matching prior. By incorporating this prior, the proposed method performs PET reconstruction via latent-space posterior sampling using the unadjusted Langevin algorithm, while enabling posterior uncertainty quantification. Numerical experiments demonstrate that the proposed method outperforms both classical reconstruction methods and state-of-the-art deep learning-based baselines. In future work, we will validate our method on in vivo data and improve its computational efficiency by developing more advanced inference algorithms.

\section{Acknowledgment}
This work was supported, in part, by the National Natural Science Foundation of China under Grant 62427807 and the Zhejiang Provincial ‘Jianbing Lingyan+X’ Science and Technology Program under Grant 2025C01127.





\bibliographystyle{IEEEtran}
\bibliography{refs}
%
%
%

\end{document}